\documentclass[aps,prb,floatfix,twocolumn,showpacs]{revtex4}
\usepackage{amsmath}
\usepackage{graphicx}
\usepackage{float}
\usepackage{amssymb}
\usepackage{times}
\DeclareGraphicsExtensions{.eps,.ps}
\begin{document}

\title{Anisotropic spin and charge excitations in superconductors:
signature of electronic nematic order}

\author{Ying-Jer Kao}
\author{Hae-Young Kee}
\affiliation{Department of Physics, University of Toronto, Toronto,
Ontario, Canada M5S 1A7}

\date{\today}

\begin{abstract}
We study spin and charge susceptibilities in the $d$-wave superconducting 
state whose underlying electronic dispersion is anisotropic
due to the formation of the electronic nematic order.
We show that the amplitude of the incommensurate peaks 
in the spin susceptibility near $(\pi,\pi)$ 
reveals a pronounced anisotropy in the momentum space. 
The relevance of our findings to the magnetic scattering pattern 
observed in recent neutron scattering measurements on untwinned 
YBa$_2$Cu$_3$O$_{6+x}$ is discussed.
In the charge channel, we identify a well-defined collective mode 
at small momentum transfer with strong anisotropic amplitude
depending on the direction of momentum transfer, which is
associated with the broken symmetry due to nematic ordering.
\end{abstract}

\maketitle

\section{Introduction}
It has been recently proposed that quantum analog of liquid crystal
states -- dubbed as electronic smectic and nematic phases --
may play an important role in physics of strongly correlated  systems.
\cite{kivelson98,kivelson03,cheong,eisenstein1,eisenstein2,eisenstein3,tranquada}
The smectic and nematic phases are characterized by the broken uni-directional
translational and rotational symmetry, and broken rotational
symmetry, respectively. When electrons reside on a lattice, the broken translational
and rotational symmetries correspond to the broken point group symmetry
of the underlying lattice. For example, the electronic nematic phase
is characterized by the broken $x-y$ symmetry on a square lattice,
and the collective modes associated with the broken symmetry.
The direct or indirect evidences of electronic smectic and nematic
phases has been reported in various strongly correlated systems,
via different experimental techniques.
\cite{cheong,eisenstein1,eisenstein2,eisenstein3,tranquada,mook00,ando}

In particular, a recent neutron scattering measurement on untwinned
YBa$_2$Cu$_3$O$_{6+x}$ have indicated the possibility of a nematic liquid phase in
high temperature superconductors.\cite{hinkov04}
Neutron scattering experiments, which provide direct information
on the spin dynamics as well as static magnetic order, have been
widely and successfully used to understand novel phenomena in high
temperature superconductors.
However, theoretical interpretations of the origin of incommensurate spin
excitations observed by neutron scattering experiments are still
controversial.\cite{mook98}  
One explanation of this phenomenon is that it
is due to Fermi surface nesting.\cite{fs}  However, it was also proposed that
this phenomena can be understood  due to the  formation of one dimensional
stripes in the system.\cite{kivelson98,mook00}  
A recent neutron experiment\cite{hinkov04} shows that the spin fluctuations in this material
are indeed of two-dimensional nature: 
the locus of maximum spin fluctuation spectral weight
approximately forms a circle in the momentum space. However,
there exists a strong anisotropy in the amplitude and the width of
the incommensurate peaks depending on the direction of the momentum transfer.
These experimental results
may indicate that the electronic structure in the high temperature
superconductors is neither 
rigid one-dimensional stripe arrays, nor an isotropic two dimensional system.

In this paper, we study incommensurate spin and charge fluctuations  
in $d$-wave superconducting state which coexists with electronic nematic phase.
We find the strong intensity anisotropy in the incommensurate spin susceptibility 
near $(\pi,\pi)$ is due to the anisotropic band structure of the nematic phase. 
The incommensurabilities and the width of the incommensurate peaks
also show anisotropy in  the momentum space. 
The qualitative results bear a resemblance to the magnetic scattering pattern 
observed in the neutron scattering measurements.\cite{hinkov04,buyers}
We also study the charge dynamics, especially focus on
the effects of coupling to the collective modes.
We find that a pronounced peak appears at moderate
energies for small momentum transfer.
This peak is an overdamped collective mode of the amplitude
fluctuations at extremely small wave-vectors and grows into a sharp mode at moderate 
finite wave-vectors. 
The intensity of  this mode also shows anisotropy in the momentum plane,
which results from the broken $x-y$ symmetry of the nematic order.  

This paper is organized as follows. In the following section,
we describe the nature of nematic order,
and introduce an effective Hamiltonian for the nematic transition within the weak
coupling theory.
We also give a brief summary  for completeness the mean field results of the
effective Hamiltonian  presented previously.\cite{kee03,khavkine04}
We present  spin and charge susceptibilities in the superconducting state
which occurs on top of the nematic phase in
Secs.~\ref{sec:chispin} and \ref{sec:chicharge}, respectively. The connection 
to experimental observations is discussed in Sec.~\ref{sec:discuss}.
The conclusion and summary are given in Sec.~\ref{sec:conclusion}.

\section{Model\label{sec:model}}






For classical liquid crystals,  the nematic order parameter is  represented
by a director, which is a quadrupolar ( rank-two symmetric traceless ) 
tensor built from the spatial directions.  
In two dimensions, it changes sign under a rotation by $\pi/2$ and 
is invariant under a rotation by $\pi$.
In similar spirit, one can construct a quadrupolar order parameter for
electronic systems using the momentum operators of electrons,
${\hat Q}_{ij} = {\hat p}_i {\hat p}_j -  \frac{1}{2} {\hat p}^2 \delta_{ij} $.
The attractive interaction between quadrupole densities can lead to
the nematic order which breaks the rotational symmetry, and 
a preferred direction for the electron momenta occurs.

Within the weak coupling theory, 
the following effective Hamiltonian of quadrupole-quadrupole
density interaction has been proposed to describe the transition between isotropic
liquid to the nematic ordered phase\cite{oganesyan01,kee03,khavkine04},
where the expectation value of quadrupole density serves as the order parameter of
the nematic phase,
\begin{equation}
\mathcal{H} =  \sum_{{\bf k}} \sum_{\sigma=\uparrow,\downarrow}
{\epsilon}_{\bf k} c^{\dagger}_{\bf k,\sigma } c^{ }_{\bf k,\sigma}+ \sum_{\bf q} F_2({\bf q})
 \left( 
\textrm{ Tr} [ {\hat Q}^{ \dagger}_{\sigma} ({\bf q})  {\hat Q}_{\sigma}({\bf q})] \right) \ ,
\end{equation}
where  ${\epsilon}_{\bf k}=-2 t (\cos k_x + \cos k_y) -4 t' \cos k_x \cos k_y-\mu$ is the 
bare electronic dispersion.  $F_2({\bf q})$ is a short-range attractive
interaction, which can be written in a general form as
\[
F_2 ({\bf q}) = -\frac{F_2}{1+\kappa F_2 q^2} \, ,
\]
where $F_2>0$. ${\hat Q}({\bf q})$ is the quadrupole density tensor,
\begin{equation}
\hat{Q}_{\sigma} ({\bf q})  =\sum_{\bf k} 
c_{{\bf k}+{\bf q} ,\sigma}^{\dagger}
\begin{pmatrix}
F (k,q) &  G (k,q) \\
G (k,q)   & F (k,q)
\end{pmatrix}   c^{ }_{{\bf k},\sigma }, 
\end{equation}
where  $ F(k,q)= [(\cos k_x-\cos k_y)+(\cos(k_x+q_x)-\cos(k_y+q_y))]/2$, and
$ G(k,q)= [(\sin k_x\sin k_y)+(\sin(k_x+q_x)\sin(k_y+q_y))]/2$. Notice that
$\hat{Q}^\dagger({\bf q})=\hat{Q}(-\mathbf{q})$, indicating that
$\hat{Q}(r)$ is Hermitian. 

The mean field solution is already presented in Ref.~\onlinecite{kee03,khavkine04},
 and we give a brief summary of the result here for completeness.
To solve the model at the mean field level, we decouple the interaction 
at $\textbf{q}\rightarrow 0$ and define two order parameters, 
\begin{eqnarray}
\Delta_N &=&F_2 \langle Q_{xx}(0) \rangle = F_2 \sum_k f(\epsilon_k) F(k),\nonumber\\
\Delta_N' &=& F_2 \langle Q_{xy}(0) \rangle =  F_2 \sum_k f(\epsilon_k) G(k), 
\label{mf}
\end{eqnarray} 
where $F(k)=(\cos k_x-\cos k_y) $ and $G(k)=\sin k_x\sin k_y$.
 The electron dispersion in the nematic phase is renormalized to
\begin{eqnarray}
\tilde\epsilon_{\bf k}  
&= & -2 t (1+\Delta_N/2t) \cos k_x -2 t (1-\Delta_N/2t) \cos k_y \nonumber \\
& &-4 t' \cos k_x \cos k_y-\Delta_N' \sin k_x \sin k_y-\mu. \label{nematicdisp}
\end{eqnarray} 
Non-zero values of  $\Delta_N$ and $\Delta_N'$ indicate the  
lattice rotational symmetry is broken and there exists a preferred direction
for the electron momentum. At the mean field level,
with a single value of $F_2$,  it is found that $\Delta_N'$ always vanishes,
which indicate the deformation of the Fermi surface is along the underlying lattice
axes. 
It is also found that $\Delta_N$ jumps at the transition 
from the isotropic phase to the nematic phase by tuning the chemical potential and/or
the strength of interaction, which leads to  
a dramatic change of the Fermi surface topology -- a closed to an open Fermi surface 
-- at the transition.
Therefore, the nematic ordered state  describes electrons
with anisotropic hopping and the topology of the Fermi surface changes dramatically
at the transition.\cite{kee03,khavkine04}


Fluctuations around the mean field states can be described by amplitude
and phase (i.e., orientational) deformations of the order parameter, and we
obtain the collective mode spectrum 
at the Gaussian level. 
In the continuum case, the phase mode is an overdamped Goldstone
mode and the amplitude mode is gapped\cite{oganesyan01}. In the lattice
case, the nematic order is pinned to the underlying lattice, so the phase
mode is also gapped. 
The form factors associated with the amplitude and phase fluctuations are
$F(k,q)$ and $G(k,q)$ respectively.\cite{footnote}

In the following sections, we study the spin and charge susceptibilities in the
$d$-wave superconducting state, which occurs on top of the
nematic liquid with anisotropic dispersion. 
The quasiparticle dispersion in the superconducting state 
is given by  $E_k=\sqrt{\tilde\epsilon_k^2+\Delta_k^2}$,
with $\tilde\epsilon_k$ is the electronic  dispersion in the nematic phase
given by Eq.~(\ref{nematicdisp}), and $\Delta_k=\Delta_{sc}(\cos k_x-\cos k_y)/2$ 
is the $d$-wave superconducting gap.


\begin{figure}[t]
\includegraphics[width=2.8in,clip]{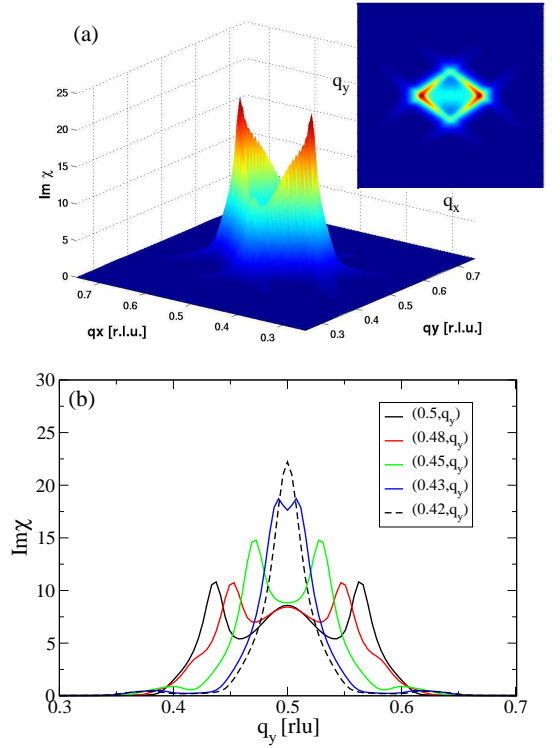}
\caption{Imaginary part of the spin susceptibility at $\omega/2t=0.12$. (a)
  3D plot of $Im \chi$ shows strong amplitude anisotropy in the $q_x-q_y$ plane near
  $(\pi,\pi)$. (Inset) 2D plot of  $Im \chi$. The incommensurability
  patten forms a ``diamond'' shape. (b) Line scans of $Im\chi$  at different
$q_x$ values. The incommensurabilities are  $\delta_x=0.08$ and $\delta_y=0.07$ r.l.u.}
\label{fig:spinsusc}
\end{figure}

\section{Dynamical Spin Susceptibility\label{sec:chispin}}
To analyze the dynamical spin susceptibility, we employ the standard random phase
approximation (RPA) at the exchange interaction and the quadrupole density
interaction.  
Spin susceptibility in RPA is given by 
\begin{equation}
\chi^s(q,\omega)=\frac{\chi^s_0}{1+J(q)\chi^s_0},
\end{equation}
where $J(q)=J (\cos q_x+\cos q_y)/2$ is the exchange interaction, and
$\chi^s_0$ is the susceptibility irreducible to $J(q)$. $\chi^s_0$ includes
the bare spin susceptibility as well as
the coupling between the spin density and the collective modes of the nematic order, 
which can be written as\cite{kadanoff64} 
\begin{equation}
\chi^s_0=\chi^s_{00}-\sum_{a=1,2}\frac{V^s_a F_2(q) \bar{V}^s_a}{1+ F_2(q) \Pi^s_a}
\equiv \chi^s_{00}+\chi^s_N,
\end{equation}
where $\chi^s_{00} \sim \langle [S_z,S_z]\rangle_0$ is the bare spin
susceptibility, $V^s_a  \sim \langle [S_z,\delta \Delta_{N,z}^a]\rangle_0$, 
$ \bar{V}^s_a \sim \langle [\delta \Delta_{N,z}^a,S_z]\rangle_0$ represents the
coupling between the spin density and the nematic order fluctuations, and 
$\Pi^s_a  \sim \langle [\delta \Delta_{N,z}^a,\delta
\Delta_{N,z}^a]\rangle_0$ is the
correlation between nematic order fluctuations. 
Here $a = 1,2$ corresponds to the amplitude and the phase mode, respectively,
and $\Delta_{N,z} = \Delta_{N,{\uparrow}} - \Delta_{N,{\downarrow}}$.
In the superconducting
state, these quantities are given by  the BCS-type expression  
assuming that the d-wave superconductivity is build on top of the nematic phase,
\begin{eqnarray}
\lefteqn{\chi^{s}_{00},V^s_a,\Pi^s_a(q,\omega)  = \sum_k M(k,q) \;\;\; \times } & & \nonumber \\
& & \left[\frac{1}{2}\left(1+\frac{\tilde\epsilon_k\tilde\epsilon_{k+q}+\Delta_k\Delta_{k+q}}
{E_kE_{k+q}}\right)\frac{f(E_{k+q})-f(E_k)}{\omega-(E_{k+q}-E_k)+i\delta} \right. \nonumber \\
& + & \frac{1}{4}\left(1-\frac{\tilde\epsilon_k\tilde\epsilon_{k+q}+\Delta_k\Delta_{k+q}}
{E_kE_{k+q}}\right)\frac{1-f(E_{k+q})-f(E_k)}{\omega+(E_{k+q}+E_k)
+i\delta} \nonumber \\
& + & \left. \frac{1}{4}\left(1-\frac{\tilde\epsilon_k\tilde\epsilon_{k+q}+\Delta_k\Delta_{k+q}}
{E_kE_{k+q}}\right)\frac{f(E_{k+q})+f(E_k)-1}{\omega-(E_{k+q}+E_k)+i\delta}\right],\nonumber\\
& & \ 
\label{chi_s}
\end{eqnarray}
where $\tilde\epsilon_k$ is the anisotropic dispersion given by Eq.~(\ref{nematicdisp}),
$\Delta_k=\Delta_{sc}(\cos k_x-\cos k_y)/2$ is the $d$-wave superconducting
gap, $E_k=\sqrt{\tilde\epsilon_k^2+\Delta_k^2}$ is the quasiparticle energies,
$f$ is the Fermi function and $M(k,q)$ are form-factors given by 
\begin{equation}
M(k,q)=
\begin{cases}
1 & \text{for \,$\chi^s_{00}$},\\
F(k,q) & \text{for \,$V^s_1$},\\
G(k,q) & \text{for \,$V^s_2$},\\
F^2(k,q) & \text{for \,$\Pi^s_1$},\\
G^2(k,q) & \text{for \,$\Pi^s_2$}.
\end{cases}
\label{formfactor}
\end{equation}

For the bare electron dispersion, we use the tight-binding parameters in
Ref.~\onlinecite{norman00}, given by $t=138$meV, $t'=-0.24 t$, and
$\mu=-0.634 t$. The superconducting gap is  $\Delta_{sc}=0.252 t\approx 35$meV.
The choice of $J=t$ gives the resonance at $(\pi,\pi)=(0.5,0.5)$ reciprocal-lattice
units (r.l.u.) at $\Omega_0=0.29t\approx 40$meV, corresponding
to the resonance peak observed in optimally doped YBa$_2$Cu$_3$O$_{6+x}$. 
We take the nematic order parameter, $\Delta_N/2t=0.1$,
temperature $T/2t=0.005$ and $\delta/2t=0.005$ in the numerical calculations. 
The momentum integrals are summed over a $512 \times 512$ mesh
in the Brillouin zone.
We set the parameters in $F_2(\textbf{q})$ to be $F_2/2t=0.49$ and $\kappa=1$. 
This set of $F_2$, $\mu$, and $\Delta_N$ satisfies the mean field
equation of Eq~(\ref{mf}).

\begin{figure}[t]
\includegraphics[width=2.5in,clip]{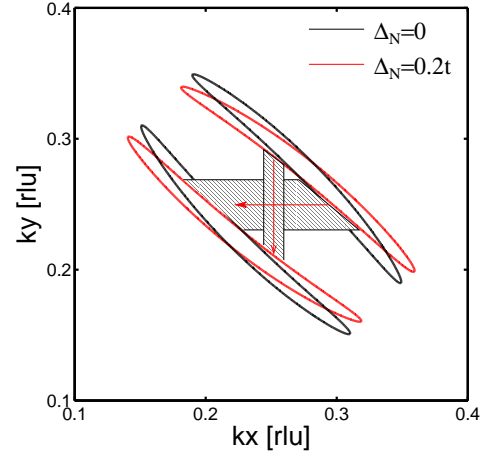}
\caption{Energy contour plots for quasiparticle dispersion
$E_k$ and $E_{k+Q}=\omega/2 = 0.12t$ 
in d-wave superconducting state. Black and red curves are
 contour with  $\Delta_N/2t=0$, and $\Delta_N/2t=0.1$, respectively. 
 Red arrows are vectors with the same
 length. In the nematic phase, rotational symmetry is broken, so the
 nesting vector for the vertical shift is no longer a nesting vector for
 the horizontal shift. And the difference in the nesting areas (shaded areas)
 results in the amplitude anisotropy. }
\label{fig:contour}
\end{figure}

Figure~\ref{fig:spinsusc} shows imaginary part of the RPA spin
susceptibility at $\omega=0.24t \approx$ 33 meV, slightly below the resonance.
A few important observations follow: 
first, the incommensurate peaks in the nematic state show clear anisotropy
along $(\pi,\pi\pm\delta_y)$ and $(\pi\pm\delta_x,\pi)$ directions. In
Fig.~\ref{fig:spinsusc}(b), we show several $q_y$ line scans at the
different fixed $q_x$ values. The amplitude of the incommensurate peaks
is small initially  along the $q_y$ direction and continues to grow until finally the
amplitude becomes maximum along the $q_x$ direction. 
Second,  the incommensurabilities  are given by  $\delta_x=0.08$ and $\delta_y=0.07$ r.l.u. 
The incommensurability pattern shows a diamond shape, 
slightly elongated along the $q_x$ direction which is the direction 
of the elongated Fermi surface due to nematic ordering 
(Fig.~\ref{fig:spinsusc}(a), inset). 
Third, there is a strong anisotropy in the width of the incommensurate peaks. 
The ratio between the full width of half maximum of the incommensurate peak 
at (0.42,0.5) and that at (0.5,0.43) is about 0.58. 
Finally, the tendency of anisotropy in the incommensurate peaks
depends on frequency. At higher frequencies, closer to the resonance mode,
the anisotropy gets weaker, and disappears at the resonance mode around 40 meV.
These results bear a resemblance to the magnetic scattering pattern
observed in recent neutron scattering experiments in YBa$_2$Cu$_3$O$_{6+x}$.\cite{hinkov04,buyers}
However, the locus of the maximum spectral weight shows approximately
circular shape in experimental data\cite{hinkov04}, 
while we find a diamond-like pattern,
which is similar to that found in Ref~\onlinecite{norman00} based on the same set of parameters 
for the bare band structure.

We find the major contribution of the anisotropy of amplitude and width of 
incommensurate peaks in the momentum space comes from the bare
spin susceptibility $\chi^s_{00}$ due to the anisotropic band structure,
while the contribution from the coupling to the collective modes,
$\chi^s_N$ is negligible, because around $\textbf{q}=(\pi,\pi)$, $V^s_a$
are small due to the form-factors. 

The anisotropy of the amplitude and
incommensurability can be understood from the quasiparticle energy
dispersion $E_k$. At low temperatures, we notice $Im\chi^s_{00}
\sim \sum_k \delta(\omega-E_{\bf  k}+E_{\mathbf{k}+\mathbf{q}})$ apart from the
coherence factors. The major contribution of the incommensurability 
is due to the nesting between the
energy contour around the node, $E_{\bf k} \sim \omega/2\approx 0.12t$, to the same contour displaced by
$\mathbf{Q}=(\pi,\pi)$, $E_{\mathbf{k}+\mathbf{Q}}\sim \omega/2$.\cite{brinckmann99} 
The best nesting vector is horizontal or vertical offset to $(\pi,\pi)$, at 
$\mathbf{q}=(\pi\pm\delta_x,\pi),(\pi,\pi\pm\delta_y) \equiv (\pi,\pi) +\delta {\bf q}$.
\cite{schulz90} In Fig.~\ref{fig:contour}, we plot the energy contours for 
$E_{\bf  k},E_{\mathbf{k}+\mathbf{Q}}=\omega/2= 0.12 t $ for the isotropic
case (black curves) and the nematic case (red curves). For the isotropic dispersion, 
the  horizontal and the vertical incommensurabilities are equal. 
On the other hand, for the anisotropic dispersion due to the nematic ordering, 
the $x-y$ rotational symmetry is broken and a preferred direction is selected. 
In Fig.~\ref{fig:contour}, the
vertical arrow corresponds to a nesting vector $\delta{\bf q} =(\delta_y,0)$, and the
horizontal arrow corresponds to the same vector, $(0,\delta_x)$ rotated by 90$^\circ$.
However, with finite $\Delta_N$, the vector $(0,\delta_x)$ is no longer a nesting vector, 
which explains the locus of maximum spin spectral weight occurs at
 slightly different incommensurate vectors, $\delta_x$ and $\delta_y$. 
The nesting regions for
the horizontal and vertical offset vectors (Fig.~\ref{fig:contour}, shaded
areas) become different due to the
nematic order, which results in the pronounced anisotropy in the amplitudes of the
incommensurate peaks.    

\begin{figure}[t]
\includegraphics[width=2.2in,clip]{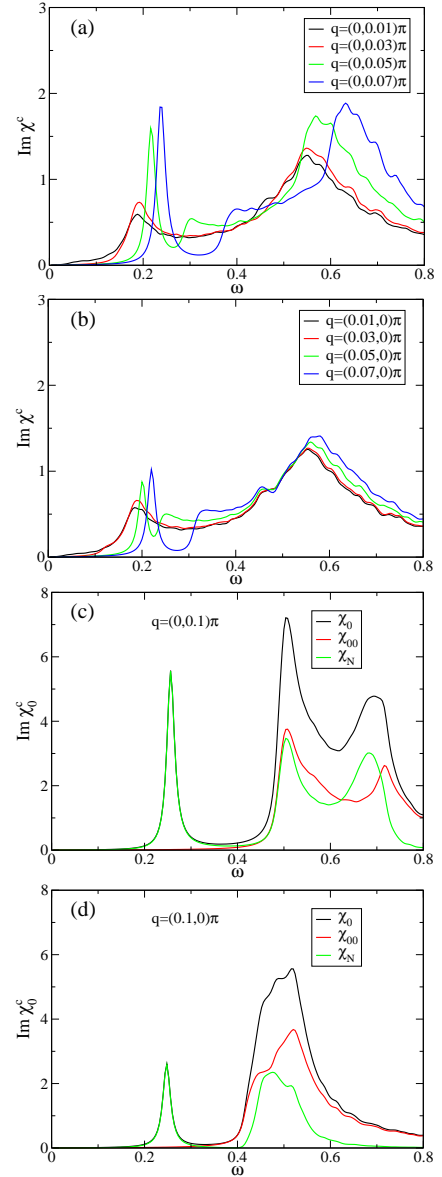}
\caption{Imaginary part of the charge susceptibilities. (a) and (b) are
  plots of RPA susceptibilities $Im \chi^c$ in $q_x$ and
  $q_y$ directions with different $|q|$ values. (c) and (d) are the 
  susceptibilities $Im \chi^c_0$ at $\mathbf{q}=(0,0.1\pi),(0.1\pi,0)$ respectively. The
  contribution can be separated into $\chi_{00}$, the susceptibility from
  the band structure and $\chi_N$, the susceptibility from coupling to the
  nematic fluctuations. $\chi_N$ gives the contribution to the collective
  mode at lower energy.  $\omega$ is in units of $t$.}
\label{fig:chargesusc}
\end{figure}

\section{Dynamical Charge Susceptibility\label{sec:chicharge}}
To analyze the dynamical charge susceptibility, we employ RPA at the quadrupole density
interaction and on-site Coulomb interaction. Similar to the previous section, the charge
susceptibility in RPA is given by 
\begin{equation}
\chi^c(q,\omega)=\frac{\chi^c_0}{1+U\chi^c_0},
\end{equation}
where $U=2t$ is the on-site Coulomb interaction, and
$\chi^s_0$ is the susceptibility irreducible to $U$.
$\chi^c_0$ includes the bare charge susceptibility $\chi_{00}^c$, as
well as the coupling between the density and the nematic order, which can be
written as follows,
\begin{equation}
\chi^c_0=\chi^c_{00}-\sum_{a=1,2}\frac{V^c_a F_2(q) \bar{V}^c_a}{1+ F_2(q) \Pi^c_a}\equiv \chi^c_{00}+\chi^c_N,
\end{equation}
where $\chi^c_{00} \sim \langle [\rho,\rho]\rangle_0$ is the bare density
susceptibility, $V^c_a  \sim \langle [\rho,\delta \Delta_N^a]\rangle_0$, 
$ \bar{V}^c_a \sim \langle [\delta \Delta_N^a,\rho]\rangle_0$ represent the
coupling between the density and the nematic order fluctuations, and 
$\Pi^c_a  \sim \langle [\delta N_a,\delta N_a]\rangle_0$ the
correlation between nematic order fluctuations. 

In the superconducting state, these quantities are given by  
the BCS-type expression  assuming that the superconductivity occurs on top of
the nematic phase,
\begin{eqnarray}
\lefteqn{\chi^{c}_{00},V^c_a,\Pi^c_a(q,\omega) = \sum_k M(k,q) \;\;\; \times} \nonumber \\
& & \left[\frac{1}{2}\left(1+\frac{\tilde\epsilon_k\tilde\epsilon_{k+q}-\Delta_k\Delta_{k+q}}
{E_kE_{k+q}}\right)\frac{f(E_{k+q})-f(E_k)}{\omega-(E_{k+q}-E_k)+i\delta} \right. \nonumber \\
& + & \frac{1}{4}\left(1-\frac{\tilde\epsilon_k\tilde\epsilon_{k+q}-\Delta_k\Delta_{k+q}}
{E_kE_{k+q}}\right)\frac{1-f(E_{k+q})-f(E_k)}{\omega+(E_{k+q}+E_k)
+i\delta} \nonumber \\
& + & \left. \frac{1}{4}\left(1-\frac{\tilde\epsilon_k\tilde\epsilon_{k+q}-\Delta_k\Delta_{k+q}}
{E_kE_{k+q}}\right)\frac{f(E_{k+q})+f(E_k)-1}{\omega-(E_{k+q}+E_k)+i\delta}\right]\nonumber\\
& & \ 
\end{eqnarray}
and the form-factors, $M(k,q)$ have the same definition as in Eq.~(\ref{formfactor}).

\begin{figure}[t]
\includegraphics[width=2.5in,clip]{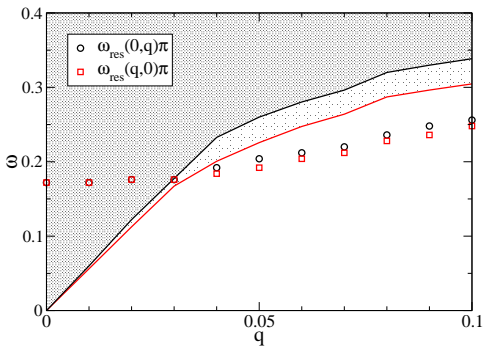}
\caption{The dispersion of the collective mode in $(0,q)$(black) and $(q,0)$(red)
  directions. Shaded areas are the continuum of the excitations. The
  boundaries corresponds to the onset of $Im \Pi^c_1$ in $(0,q)$(black) and
  $(q,0)$(red) directions. $\omega$ is in units of $t$.}
\label{fig:dispersion}
\end{figure}

We show in Fig.~\ref{fig:chargesusc} the imaginary part of the RPA charge
susceptibilities $\chi^c$ in the $(q_x,0)$ and $(0,q_y)$ directions at
small momentum transfer $\mathbf{q}$. There is a peak at low energy, and a
broader spectrum at higher energies. We show  the momentum dependence of
$Im \chi^c$ along the $(q_x,0)$ and $(0,q_y)$ directions
in Fig.~\ref{fig:chargesusc}(a) and (b). We find the peaks become narrower with
the increase of $|\mathbf{q}|$, and and a sharper peak emerges at around
$|\mathbf{q}|\approx 0.04\pi$.  We see there is a clear difference for
$Im \chi^c$ in the two directions of $(q_x,0)$ and $(0,q_y)$, which indicates
the broken lattice rotational symmetry due to the nematic order. 
To analyze the origin of the low energy peak, 
we plot in Fig.~\ref{fig:chargesusc}(c) and (d) $\chi_0$, $\chi_{00}$ and $\chi_N$ separately at
$\mathbf{q}=(0,0.1\pi), (0.1\pi,0)$.  The bare charge susceptibility,
$\chi_{00}$ contributes mainly to the broader spectrum at  high
energies with different spectral weights for $(0,0.1 \pi)$ and $(0.1\pi,0)$.
The coupling of the density to the nematic order fluctuations 
provides a well-defined resonance peak at low energy, in addition to the
high energy spectrum. This corresponds to the amplitude mode of the nematic fluctuations.
Comparing Fig.~\ref{fig:chargesusc}(c) and
(d), the anisotropy between the wave-vectors, $(0,0.1\pi)$ and $(0.1\pi,0)$
is manifested in the amplitudes of the resonance peak
and the weight and shape of the high energy spectrum. 

To understand the dispersion relation of the amplitude mode,
we summarize the dispersion of the mode in Fig.~\ref{fig:dispersion}.
The resonance frequency $\omega_{\text{res}}$ is given by the peak
position of $Im \chi^c$. The shaded areas correspond to a continuum of
excitations and the boundaries are given by the onset of $Im
\Pi^c_1(\mathbf{q},\omega)$. We find that the collective mode is overdamped in
the small-$q$ region, and becomes a well-defined mode at moderate $q$. We
also see that when the mode becomes sharp, there exists a weak anisotropy in the
resonance frequencies in the two different directions of momentum transfer, $|\mathbf{q}|$.  
This sharp mode is the collective mode associated with the nematic order,
so it can provide a direct observation for the nematic order in the real materials.
\cite{yamase04}




\section{Discussion\label{sec:discuss}}
The incommensurate spin susceptibility obtained from our calculation based
on the idea of coexisting nematic and superconducting orders shows
great similarities to the experimental magnetic scattering pattern
reported in a recent inelastic neutron experiment\cite{hinkov04}: 
while the basic character of spin excitations is two-dimensional,
the amplitude and the width of the incommensurate peaks along the locus
of maximum spin spectral weight is modulated in a one-dimensional fashion.
We  find that at constant frequency scan,
the amplitude of the spectral weight is maximum at incommensurate 
wave-vector along $q_x$-direction (or $q_y$, depending on the direction of
the Fermi surface deformation due to the nematic order), 
and it continues to decrease
to the wave-vector along $q_y$($q_x$)-direction, as shown in Fig.~\ref{fig:spinsusc}.
However, the observed incommensurate fluctuations  approximately form a
circle (within the numerical convolution error bars)
in the momentum space, while our theoretical calculation shows
incommensurate pattern of a diamond shape. This diamond shape feature is 
common to the theoretical models incorporating the RPA method.\cite{brinckmann99,norman00,kao00} 
The anisotropy of the spectral weight depending on the direction in 
the momentum space originates from the anisotropic band structure of the electrons 
due to the formation of the nematic order. 
The contribution from the coupling to the nematic order fluctuations for a large 
momentum transfer near $(\pi,\pi)$ is negligible. 

On the other hand, a collective mode at moderate energies for small momentum transfer
${\bf q}$, is observed in the density fluctuations.
Due to the opening of the superconducting gap, the space of
particle-hole continuum at low energy is limited for nodal excitations.
Therefore, the collective mode becomes sharp around $|q|\approx 0.04\pi$,
while it is  overdamped as $q\approx 0$ shown in Fig.~\ref{fig:chargesusc}. 
The amplitude of this mode shows a strong anisotropy depending on the direction
of  the momentum transfer, reflecting the $x-y$ symmetry breaking nature of the nematic order, 
but its dispersion depends weakly on the direction of the 
momentum transfer. 
Since this is a well-defined mode and
well separated from the high energy contributions from the band structure,
it can be  detected by the inelastic X-ray experiments or electron scattering. 
It also leads to the additional signal in the optical conductivity at 
the frequency of collective mode.

Another implication of the nematic ordering is the distortion of the underlying
Fermi surface due to the formation of the nematic order.
Angle resolved photoemission
spectroscopy shows strong in-plane anisotropy in the electronic structure,
as well as in the magnitude of the superconducting gap.\cite{lu}
Our results suggest that the in-plane anisotropy of spin dynamics
originates from the anisotropy of the hopping integrals.
While  the nematic order induces the hopping integral anisotropy 
of $t_x/t_y \sim 1.22$ with the set of parameters we used,
we cannot exclude another possible explanation of the orthorhombicity of
CuO$_2$ plane\cite{manske}  
and its connection to the CuO chain in YBa$_2$Cu$_3$O$_{6+x}$.
According to the band theory\cite{andersen} the ratio
between the hopping integrals along the $x$ and $y$ directions scales roughly
as $(b/a)^4$ which is at most $1.06$ around optimal doping.
We find that this ratio of $t_x/t_y = 1.06$ for only the nearest neighbor hopping
integral can not produce the visible anisotropy
in the spin dynamics of the $d$-wave superconductor within RPA :
$t_x/t_y =1.06$ leads to a ratio of $1.1$ in the peak amplitudes
between $(0.42,0.5)$ and $(0.5,0.43)$, while our set of parameters leads to
the ratio of $2.3$, which depends on the strength of $F_2$ and $\mu$.

However, this is only a quantitative argument, and
further studies, such as identification of the collective modes, are required
to clarify the origin of anisotropy observed in the spin excitations.
It would be also interesting to study the doping dependence of 
the strength of anisotropy in the spin excitations.
While the structural anisotropy (the ratio 
between the lattice constants, $b$ and $a$) decreases as oxygen content
decreases\cite{jorgensen,casalta}, the tendency toward nematic ordering gets
stronger in the underdoped regime.\cite{kivelson03}
Therefore, the strength of anisotropy in the magnetic scattering signal
will either  decrease or increase, depending on the origin
of anisotropic hopping integrals: it 
gets stronger (weaker) if its origin is due to the nematic order
(structural anisotropy), as the hole doping concentration decreases.
While the existing data with different doping concentrations\cite{hinkov04,
buyers} indicates a stronger anisotropy in underdoped YBCO,
a systematic study is necessary to determine the origin.
Another clue may come from the study on the temperature dependence of
the strength of anisotropy in the magnetic scattering pattern. 
Since the structural anisotropy does not depend 
on temperature (below 700K),\cite{jorgensen}
the strength of the anisotropy due to orthorhombicity 
depends weakly on temperature, only via thermal broadening and 
temperature dependence of the superconducting order parameter.
On the other hand,  the anisotropy  induced by the nematic order  
gets weaker as temperature increases,\cite{kivelson03,khavkine04,buyers} and
eventually disappears at the nematic-isotropic transition temperature.

\section{Summary and Conclusion \label{sec:conclusion}}

While the crystalline stripes have been observed directly in maganese oxide
compounds via X-ray diffraction\cite{cheong}, 
the nematic phase, which can be viewed as strongly fluctuating stripes,
requires further theoretical and experimental studies 
to make direct connections to  real materials such as high temperature superconductors.\cite{kivelson03}
Here we have studied  spin and charge susceptibilities in a superconducting phase
whose underlying electronic dispersion is anisotropic due to
the nematic order, and aimed to understand the effects of the nematic order
on spin and charge excitations at the level of weak coupling theory. 

We consider a phase
where $d$-wave superconducting order, $\Delta_k$, occurs on top of the electronic nematic phase.
The underlying electronic dispersion, $\tilde\epsilon_k$ breaks the $x-y$ symmetry,
and the banana shape of energy contour for
the superconducting quasiparticle $E_k =\sqrt{\tilde\epsilon_k^2+\Delta_k^2}$
is tilted due  to the formation of the nematic order, $\Delta_N$ 
($\tilde\epsilon_k = \epsilon_k + \Delta_N (\cos(k_x)-\cos(k_y)$).
The standard RPA is used to compute the spin and charge susceptibilities,
where the irreducible susceptibility includes not only the bare one
but also the coupling to the collective modes associated with the broken
$x-y$ symmetry in the nematic phase.

Motivated by a recent neutron scattering experiment on detwinned optimally doped
YBa$_2$Cu$_3$O$_{6+x}$\cite{hinkov04}, we study the spin excitations near the resonance mode,
$(\pi,\pi)$, and find that the amplitude and width of the incommensurate peaks,
which appear at a set of incommensurate wave vectors, $|{\bf q}|$,
at constant frequency below the resonance, depends on
the direction of the momentum transfer. 
For example, a pronounced anisotropy in the amplitude of incommensurate
peaks between $(\pi,\pi \pm q_y)$ and $(\pi \pm q_x,\pi)$ where $q_x \approx q_y$
has been found, which shows close similarities to the magnetic scattering pattern
observed in the neutron scattering experiment.
Our analysis shows that the main contribution of this anisotropy stems from
the anisotropic band structure due to the nematic order.
However, to clarify the origin of the anisotropic band dispersion,
further studies, such as temperature and doping dependence
of the anisotropy in the signal strength, are required. 

In the charge channel, the coupling between the density and the collective modes
shows a well-defined collective mode for small momentum transfer.
While this mode is overdamped due to the particle-hole continuum 
near $q\approx 0$, it becomes sharp as $q$ increases, due to fact that the
space of Landau damping is restricted to the small momentum for low energy
because of the opening of the superconducting gap.
This mode can be detected by inelastic X-ray scattering, and may lead to
additional signal in optical conductivity.

It is interesting to note that the form-factors for  the nematic fluctuations 
are very similar to those of the Raman vertices for the $B_{1g}$ and $B_{2g}$ channels, 
and thus the collective mode in the charge susceptibility will produce
an additional Raman response in the particular channels.
Also the coupling of the nematic fluctuations to the lattice degrees of freedom may result in some
anomalous phonon responses. These are interesting questions and 
subjects of future studies. 

\begin{acknowledgments}
We thank Steven A. Kivelson, B. Buyers, and B. Keimer
 for valuable discussions. We also thank
K. Sengupta for critical reading of the manuscript.
This work was supported by NSERC of Canada, Canada Research
Chair, Canadian Institute for Advanced Research, Alfred P.~Sloan 
Research Fellowship (HYK), and Emerging Material Knowledge program
funded by Materials and Manufacturing Ontario (YJK, HYK). 
\end{acknowledgments}


\begin{thebibliography}{28}
\bibitem[{foo()}]{footnote}
\bibinfo{note}{In principle, one should diagonalize a two by two matrix to
  obtain the collective mode spectrum; however, the off-diagonal mixing terms
  are small in this model, so we only need to consider the diagonal terms.}
\end{thebibliography}

\end{document}